\documentclass[pre,aps,twocolumn,superscriptaddress]{revtex4}

\usepackage[dvips]{graphicx}
\usepackage{amssymb,amsfonts,amsmath}
\usepackage{color}
\usepackage{ulem}

\newcommand{\rv}{{\mathbf r}}

\newcommand{\J}{{\bf J}}

\newcommand{\F}{{\bf F}}

\newcommand{\Xv}{{\bf X}}

\newcommand{\vel}{{\bf v}}

\newcommand{\chib}{{\boldsymbol \chi}}

\newcommand{\myint}{\int}

\newcommand{\rnt}{{\bf r}^N\!\!,t}
\newcommand{\kT}{k_{\rm B}T}

\begin{document}

\title{Power functional theory for Brownian dynamics}

\author{Matthias Schmidt}
\affiliation{Theoretische Physik II, Physikalisches Institut, 
  Universit{\"a}t Bayreuth, D-95440 Bayreuth, Germany}

\author{Joseph M. Brader$^*$}
\affiliation{Soft Matter Theory,
  University of Fribourg, CH-1700 Fribourg, Switzerland}
\email{joseph.brader@unifr.ch}

\begin{abstract}
Classical density functional theory (DFT) provides an exact
variational framework for determining the equilibrium properties of
inhomogeneous fluids.  We report a generalization of DFT to treat the
non-equilibrium dynamics of classical many-body systems subject to
Brownian dynamics.  Our approach is based upon a dynamical functional
consisting of reversible free energy changes and irreversible power
dissipation. Minimization of this `free power' functional with respect
to the microscopic one-body current yields a closed equation of
motion.  In the equililibrium limit the theory recovers the standard
variational principle of DFT. The adiabatic dynamical density
functional theory is obtained when approximating the power dissipation
functional by that of an ideal gas.  Approximations to the excess
(over ideal) power dissipation yield numerically tractable equations
of motion beyond the adiabatic approximation, opening the door to the
systematic study of systems far from equilibrium.
\end{abstract}

\date{Received 25 April 2013; J. Chem. Phys. {\bf 138}, 214101 (2013);
doi:\ 10.1063/1.4807586}

\maketitle

\section{Introduction}
The dynamics of soft matter is a topic of considerable
current interest from both an experimental and theoretical perspective
\cite{dhont_book,wagner_book,brader10review}. Among the various
theories aiming to provide a microscopic description of complex
transport and pattern formation phenomena, the dynamical density
functional theory (DDFT) \cite{evans79,marinibettolomarconi99} has
emerged as a prime candidate.  Much of the appeal of DDFT arises from
its ease of implementation and its close connection to equilibrium
density functional theory (DFT), which is a virtually unchallenged
framework for the study of equilibrium properties in inhomogeneous
liquids \cite{evans79}.  The DDFT was originally suggested on the
basis of phenomenological reasoning \cite{evans79}, and has since been
rederived from the Langevin equation \cite{marinibettolomarconi99},
using projection operators \cite{espanol} and via coarse graining the
many-body Smoluchowski equation \cite{archer04ddft}.  Despite the deep
insight provided by these alternative derivations of the DDFT, the
final equation of motion for the density remains unchanged and no
guidance is provided for making a systematic, or indeed any
improvement that goes beyond the standard formulation.  A striking and
undesirable feature of DDFT is the clear asymmetry between the
treatment of spatial and temporal degrees of freedom.  The intricate
nonlocal spatial structure is in stark contrast to the simple
time-local treatment of the dynamics and suggests much potential for
improvement, particularly in view of the sophistication of temporally
nonlocal memory-function approaches such as the mode-coupling
theory~\cite{sjoegren}.

At the heart of DDFT lies the `adiabatic approximation', in which the
non-equilibrium pair correlations of the real system are approximated
by those of a fictitious equilibrium, whose density distribution is
given by the instantaneous density of the non-equilibrium system. 
Although this provides a reasonable account of
relaxational dynamics in colloidal liquids at low and intermediate
volume fraction, it does not capture the physics of the glass
transition and strongly underestimates the structural relaxation
timescale at high volume fraction \cite{reinhardt_shear}.  Moreover,
the adiabatic assumption fails when applied to even the simplest
driven systems, due to the neglect of non-affine particle motion which
generates a nontrivial current in directions orthogonal to local shear
flows \cite{brader11shear}.  These omissions make the theory incapable
of describing either long-range correlations or symmetry breaking
induced by the flow, thus putting out of reach many technologically
relevant and fundamental non-equilibrium problems, such as
shear-induced ordering \cite{wagner_book,ordering1,ordering2} or
migration effects in nonuniform channel flows \cite{migration}.

In this paper we develop a variational approach to colloidal dynamics
based upon the current, rather than the density, as the fundamental
variable.  We thus provide a natural generalization of classical DFT
to treat non-equilibrium situations.  Our framework transcends the
adiabatic approximation and provides a physically intuitive method to
go beyond DDFT.  

\section{Theory}
\subsection{Microscopic dynamics}
The overdamped dynamics of
colloidal particles can be described on a microscopic level by a
stochastic Langevin equation, based upon the assumption that the
momentum degrees of freedom equilibrate much faster than the particle
positions \cite{riskin}.  The velocity of particle $i$ at
time $t$ depends on the configuration, $\{ \rv_1,\cdots,\rv_N
\}\equiv\rv^N$, and is determined by the simple equation of motion
\begin{align}
  \gamma \vel^{\rm Lan}_i(\rnt) = \F_i(\rnt)  + \chib_i(t),
  \label{EQlangevin}
\end{align}
where $\gamma$ is a friction constant related to the bare diffusion
coefficient, $D_0$, according to $\gamma=\kT/D_0$, with $k_{\rm B}$
the Boltzmann constant and $T$ the temperature.  The stochostic,
velocity-dependent viscous force, $\gamma\vel^{\rm Lan}_i$, is
balanced by both a random force, $\chib_i(t)$, representing the
thermal motion of the solvent, and the force $\F_i$, arising from
interparticle interactions and external fields.  Due to the
fluctuation-dissipation theorem the stochastic force is
auto-correlated according to $\left\langle\chib_i(t)\chib_j(t')
\right\rangle = 2\kT\gamma\delta_{ij}\delta(t-t'){\bf 1}$. The
deterministic force acting on particle $i$ is given by
\begin{align}
  \F_i(\rnt) = -\nabla_i  U(\rv^N) -\nabla_i V_{\rm ext}(\rv_i,t) 
   + \Xv(\rv_i,t).
\label{EQconfig_force}   
\end{align}
This force depends upon the positions of all particles and consists of
three physically distinct contributions: (i) Forces arising from the
total interaction potential, $U(\rv^N)$. (ii) Conservative forces
generated by an external potential, $V_{\rm ext}(\rv,t)$. (iii)
Non-conservative forces $\Xv(\rv,t)$. Our choice to distinguish
conservative from nonconservative forces is not essential, but is made
for later convenience when developing the variational approach to the
many-body dynamics.  The noise term in \eqref{EQlangevin} is additive,
from which follows that the particle positions, $\rv_i(t)$, obtained
by integrating the velocities, are insensitive to the choice of
integration scheme employed: Ito and Stratonovich calculus both yield
the same result \cite{riskin}.  It should also be noted from
\eqref{EQlangevin} that the velocities at any time are well defined
quantities, completely determined by the stochastic force and the
instantaneous configuration of particles.

An alternative, but entirely equivalent description of Brownian
dynamics is provided by the probability density for finding the
particles in a given configuration at time $t$, which we denote by
$\Psi(\rnt)$.  The time evolution of this probability is given exactly
by the Fokker-Planck equation corresponding to \eqref{EQlangevin},
which takes the form of the many-body continuity equation
\begin{align}
  \frac{\partial \Psi(\rnt)}{\partial t} = 
  -\sum_{i}\nabla_i\cdot\J_i (\rnt),
\label{EQsmol}
\end{align}
where the sum is taken over all particles and the current of particle $i$ is
given by
\begin{align}
\J_i (\rnt) = \gamma^{-1}\Psi(\rnt) \left[ 
  \F_i(\rnt) - \kT\nabla_i \ln \Psi(\rnt)\right].  
\label{EQsmol_current} 
\end{align}
Equations \eqref{EQsmol} and \eqref{EQsmol_current} constitute a many-body
drift-diffusion equation, generally referred to as the Smoluchowski
equation \cite{dhont_book}.  The factor in square brackets appearing
in \eqref{EQsmol_current} is the total force acting on particle~$i$,
which we denote by
\begin{align}
  \F_i^{\rm tot}(\rnt)\equiv{\F}_i(\rnt) -\kT \nabla_i \ln \Psi(\rnt),
\label{EQFtot}
\end{align} 
and which includes both direct forces, as well as statistical, thermal
forces.  The (deterministic) velocity of particle $i$ can be
identified from \eqref{EQsmol_current} as
\begin{align}
  \vel_i (\rnt) = \gamma^{-1} \F_i^{\rm tot}(\rnt),
\label{EQsmol_velocity}
\end{align}
such that the current is simply $\J_i (\rnt)=\vel_i (\rnt)\Psi(\rnt)$.
For calculation of average quantities, rather than individual
stochastic particle trajectories, the information provided by the
deterministic function $\vel_i (\rnt)$ is equivalent to that of the
stochastic Langevin velocity, $\vel^{\rm Lan}_i (\rnt)$, appearing in
\eqref{EQlangevin}.

\subsection{\bf Central variables}  
An important average quantity
characterizing non-equilibrium particle dynamics is the time-dependent
one-body density,
\begin{equation}
  \rho(\rv,t) = 
  \left\langle\sum_i\delta(\rv-\rv_i(t))\right\rangle,
  \label{EQdensityDefinition}
\end{equation}
where $\rv_i(t)$ is the position of particle $i$ at time $t$,
$\delta(\cdot)$ is the Dirac distribution, and the sum is over all $N$
particles.  Within the Langevin description the angle brackets
$\langle\cdot\rangle$ indicate an average taken over many solutions of
\eqref{EQlangevin} generated using independent realizations of the
random forces, $\chib_i(t)$, and an average over initial conditions.
Within the probabilistic Smoluchowski picture the angle brackets
should be interpreted as a configurational average, to be calculated
using the probability density function $\Psi(\rnt)$ obtained from
solution of \eqref{EQsmol}. An arbitrary function of the particle
coordinates, $\hat f(\rnt)$, thus has average value $f(t)=\langle \hat
f(\rnt) \rangle\equiv\int d\rv^N\Psi(\rnt)\hat f(\rnt)$.

Within the standard DDFT the one-body density acts as the central
variable.  However, the motion of the system is better characterized
by a vector field, namely the time- and space-resolved one-body
current
\begin{align}
  \J(\rv,t) &= 
  \left\langle\sum_i\vel_i(\rnt)\,\delta(\rv-\rv_i(t))\right\rangle,
  \label{EQcurrentDefinition}
\end{align} 
where $\vel_i(\rnt)$ is given by \eqref{EQsmol_velocity}, and the angle
brackets represent a configurational average. Of course, the same
result is obtained from \eqref{EQcurrentDefinition}, if one replaces
$\vel_i(\rnt)$ with $\vel^{\rm Lan}_i(\rnt)$ and employs an average over
stochastic realizations and initial conditions. 
It is evident that the vector current
\eqref{EQcurrentDefinition} provides a more appropriate starting point
than the scalar density \eqref{EQdensityDefinition} for a theory
aiming to describe flow in a complex liquid; there are many relevant
flows for which the density is trivial, $\rho(\rv,t)=\rm const$, but
the current is not.  The same conclusion has also been arrived at
within the quantum DFT community, where time-dependent problems are
generally addressed using approaches for which the current is regarded
as the fundamental variable \cite{runge_gross}.  Once the
current is known, the density can be calculated from the continuity
equation, given in integral form by
\begin{align}
  \rho(\rv,t) &= \rho_0(\rv)-\int_{t_0}^t dt' \nabla\cdot\J(\rv,t'),
  \label{EQcontinuityEquation}
\end{align}
and expressing the local conservation of particle number.  At the
initial time $t_0$ we take the system to be in equilibrium, possibly
in an inhomogeneous state, $\rho_0(\rv)\neq \rm const$, which is
generated by the action of an external potential $V_{{\rm
    ext},0}(\rv)$. The many-body distribution $\Psi_0(\rv^N)$ at time
$t_0$ is then of Boltzmann form.  Recall that equilibrium DFT
\cite{evans79} provides a framework for calculating $\rho_0(\rv)$ for
given $V_{{\rm ext},0}(\rv)$ and interparticle interactions
$U(\rv^N)$; a very brief synopsis of DFT is given below.

\subsection{\bf Variational principle}  
In classical mechanics,
dissipative effects are commonly dealt with via Rayleigh's dissipation
function, from which frictional forces are generated by
differentiation with respect to particle velocities \cite{goldstein}.
For the Brownian dynamics under consideration, we formulate the
many-body problem in terms of a generating function, consisting of
dissipative and total force contributions as well as non-mechanical
contributions due to temporal changes in the external potential,
\begin{align}
   \hat R(\rv^N\!\!,\tilde{\vel}^N\!\!,t) =&
   \sum_i \left(\frac{\gamma}{2}
   \tilde{\vel}_i(\rnt)-\F_i^{\rm tot}(\rnt) \right)\cdot 
   \tilde{\vel}_i(\rnt)
   \notag\\&
   +\sum_i \dot V_{\rm ext}(\rv_i,t),
   \label{EQgeneratingFunction}
\end{align}
where $\dot V_{\rm ext}(\rv,t)=\partial V_{\rm ext}(\rv,t)/\partial
t$, and the $\tilde{\vel}_i(\rnt)$ are trial functions which may be
varied independently of the distribution function contained within the
thermal force term in \eqref{EQgeneratingFunction}. When averaged over
particle configurations the many-body function $\hat R$ becomes a {\it
  functional} of both the trial velocity and the distribution function
\begin{align}
  R_t[\Psi,\tilde\vel^N]  \equiv 
  \myint d\rv^N \Psi(\rnt)\hat R(\rv^N,\tilde{\vel}^N,t),
  \label{EQaverage_generator} 
\end{align} 
which is valid for an arbitrary $\Psi(\rnt)$. For notational
convenience we indicate the dependence of a functional on time $t$
using a subscript.  The functional $R_t[\Psi,\vel^N]$ is minimized by
setting equal to zero the functional derivative with respect to the
trial field,
\begin{align}
  \frac{\delta R_t[\Psi,\tilde{\vel}^N]}{\delta \tilde{\vel}_i(\rnt)}=0,
  \label{EQvariation_i}
\end{align} 
where the variation is performed at fixed $\Psi(\rnt)$ and at fixed
time $t$. From \eqref{EQvariation_i} follows directly, observing the
simple quadratic structure of \eqref{EQgeneratingFunction}, that the
minimal, physically realized trial fields are given by
$\tilde{\vel}_i=\vel_i$ and are thus related to the forces according
to \eqref{EQsmol_velocity}.  Imposing the condition that probability
is conserved throughout the dynamics, i.e.\ the many-body continuity
equation, then recovers the Smoluchowski equation, \eqref{EQsmol} and
\eqref{EQsmol_current}. When evaluated at the physical solution, the
functional becomes, upon inserting \eqref{EQsmol_velocity} into
\eqref{EQgeneratingFunction},
\begin{align}
   R_t[\Psi,\vel^N] =& -\frac{1}{2}\int d\rv^N \Psi(\rnt)
  \sum_i \F_i^{\rm tot}(\rnt)\cdot\vel_i(\rv,t)
\notag\\ & 
\quad + \int d\rv^N\Psi(\rnt)\sum_i \dot V_{\rm ext}(\rv_i,t)
\end{align}
where the first term is $-1/2$ times the total power that the system
handles at time $t$ and the second term is the non-mechanical rate of
external potential increase.  In equilibrium the velocities are zero
and the minimization condition \eqref{EQvariation_i} reduces to the
simple expression $U_N(\rv^N)+\sum_i V_{\rm
  ext}(\rv_i)-\kT\ln\Psi(\rv^N)={\rm const}$, thus recovering the
Boltzmann distribution.

We now proceed to exploit the results described above in order to
arrive at a more useful variational scheme on the one-body, rather
than the many-body, level.  In the original Hohenberg-Kohn formulation
of DFT \cite{hohenberg} the variational principle relies on the
condition that a given one-body density is generated by some external
potential (a requirement known as $v$-representability, where $v$
indicates an external potential).  The alternative `constrained
search' formulation provided by Levy \cite{levy} has the advantage
that it relies on a weaker $N$-representability condition, which, for
classical systems, requires only that the given one-body density be
generated by some many-body probability distribution \cite{brams}.
The strength of the Levy method, which we now exploit, is that it can
be readily applied out-of-equilibrium: the distribution function does
not have to be of Boltzmann form. We thus define our central
functional, which we henceforth refer to as the free power functional,
as a constrained minimization
\begin{align}
  {\cal R}_t[\rho,\J]=\min_{\tilde{\vel}^N\rightarrow \rho, \J}
  R_t[\Psi,\tilde{\vel}^N].
\label{EQlevy}
\end{align} 
Here the minimization searches at time $t$ over all possible trial
velocities which yield a desired target one-body density and target
one-body current, defined respectively by the configurational averages
\begin{align}
  \rho(\rv,t)&=\int d\rv^N\tilde\Psi(\rnt)\sum_i \delta(\rv-\rv_i),
  \label{EQav_density}\\
  \J(\rv,t)&=
  \int d\rv^N\tilde\Psi(\rnt)\sum_i \tilde{\vel}_i(\rnt)\delta(\rv-\rv_i),
  \label{EQav_current}
\end{align}
and selects the set of trial velocities $\tilde\vel_{i,\min}(\rv,t)$
which minimize $R_t$. We take the trial velocities
$\tilde\vel_i(\rv,t)$ to be parametrized by a trial distribution
$\tilde\Psi(\rnt)$, which is normalized according to $\int
d\rv^N\tilde\Psi(\rv,t)=1$, and which generates instantaneously the
velocities via
\begin{align}
  \tilde \vel_i(\rnt) = \gamma^{-1}\left(
  \F_i(\rnt)-\kT \nabla_i \ln \tilde\Psi(\rnt)
  \right).
  \label{EQtotalPowerTimesMinusOneHalf}
\end{align}
The minimization \eqref{EQlevy} hence becomes a minimization with
respect to $\tilde\Psi(\rnt)$ under the contraints
\eqref{EQav_density} and \eqref{EQav_current}. We denote the
distribution at the minimum by $\tilde\Psi_{\min}(\rnt)$.

We can now eliminate the dependence on the many-body distribution,
$\Psi(\rnt)$, appearing on the right hand side of \eqref{EQlevy}, by
requiring it to satisfy the many-body continuity equation, using the
trial velocites $\tilde\vel_{i,\min}(\rnt)$ and corresponding
distribution $\tilde\Psi_{\min}(\rnt)$ as input, i.e.\
\begin{align}
  \Psi(\rv,t) = \Psi_0(\rv^N)-\int_{t_0}^t dt' \sum_i
  \nabla_i\cdot \tilde\vel_{i,\min}(\rnt') \tilde\Psi_{\min}(\rv,t').
  \label{EQmanyBodyContinuityForTrialFields}
\end{align}
The substitution of \eqref{EQmanyBodyContinuityForTrialFields} into
\eqref{EQlevy} is consistent with minimization with respect to the
trial velocities at fixed $\Psi(\rv,t)$, because the minimization is
performed at the fixed time $t$.  The high-dimensional variation
\eqref{EQvariation_i} hence becomes the simpler one-body variational
principle
\begin{align}
  \frac{\delta {\cal R}_t[\rho,\J]}{\delta \J(\rv,t)}=0,
\label{EQvariationalPrinciple}
\end{align}
where $\rho(\rv,t)$ and the history of $\rho(\rv,t')$ and $\J(\rv,t')$
at times $t'<t$ are held constant.  Within the space of all
density and current fields, the physically realized fields
are given by the minimum condition \eqref{EQvariationalPrinciple}.
The functional ${\cal R}_t[\rho,\J]$ will depend in general on the
density and current in a complicated way, nonlocal in space and
containing memory effects in time.  The spatial nonlocality of ${\cal
  R}_t[\rho,\J]$ follows from the presence of a finite-range
interaction potential $U(\rv^N)$ in \eqref{EQaverage_generator},
whereas temporal nonlocality is a consequence of the time integral in
\eqref{EQmanyBodyContinuityForTrialFields}, which couples to the trial
velocity fields, and hence the one-body fields via the constraints, at
earlier points in time.  

\subsection{\bf Relationship to Mermin's functional} 
It is instructive
to relate the configurational contribution in
\eqref{EQaverage_generator} to the form of the many-body functional
\cite{mermin} originally introduced to treat quantum systems at finite
temperature. The generating functional can be decomposed as
\begin{align}
  R_t[\Psi,\vel^N] = & \; \dot \Omega_{\rm M}[\Psi] +
  \frac{\gamma}{2}\int d\rv \Psi(\rv,t) \vel_i^2(\rv,t)
  \notag\\
 &  -\int d\rv^N\Psi(\rnt)\sum_i
 \vel_i(\rnt)\cdot\Xv(\rv_i,t),
  \label{EQRdecomposedIntoMerminDerivative}
 \end{align}
where $\dot\Omega_{\rm M}[\Psi]$ is the total time derivative of
\begin{align}
  \Omega_{\rm M}[\Psi]=&\int d\rv^N \Psi(\rnt)
  \Big[U(\rv^N) +
\notag\\ & \quad
  \sum_i \left(V_{\rm ext}(\rv_i,t)-\mu\right)
  + \kT\ln\Psi(\rnt)
  \Big].
  \label{EQomegaMermin}
\end{align}
Apart from the absence of a kinetic term, not required for Brownian
dynamics, and the addition of time arguments, \eqref{EQomegaMermin} is
the same as Mermin's functional for the time-independent
case~\cite{mermin}; here $\mu=\rm const$ is the chemical potential.
Equation \eqref{EQRdecomposedIntoMerminDerivative} is obtained in a
straightforward way from \eqref{EQaverage_generator} via integration
by parts in space, using the many-body continuity equation in order to
make the replacement $-\sum_i\nabla_i\!\cdot\!\vel_i\Psi=\dot \Psi$, and
observing the normalization $\int \! d\rv^N\Psi=1$.

For completeness, we recall that in equilibrium the grand potential density
functional $\Omega[\rho]$ follows from \eqref{EQomegaMermin} via a
constrained Levy search in the function space of $\Psi(\rv^N)$,
\begin{align}
  \Omega[\rho] = \min_{\Psi\to\rho} \Omega_{\rm M}[\Psi]
  \label{EQomegaDensityFunctional}
\end{align}
under the contraint that $\rho(\rv)=\int
d\rv^N\Psi(\rv)\sum_i\delta(\rv-\rv_i)$ is held fixed \cite{brams}.
The structure of \eqref{EQomegaMermin} allows one to split
\eqref{EQomegaDensityFunctional} into intrinsic and external
contributions,
\begin{align}
  \Omega[\rho] = F[\rho] +
  \int d\rv \rho(\rv)(V_{\rm ext}(\rv)-\mu),
\end{align}
where $F[\rho]$ is the intrinsic free energy density functional, which
is independent of $V_{\rm ext}(\rv)$. The minimization principle
\cite{evans79} states that
\begin{align}
  \frac{\delta \Omega[\rho]}{\delta \rho(\rv)}= 0,
  \label{EQminimizationPrincipleDFT}
\end{align}
where the left hand side can be written as
$\delta\Omega[\rho]/\delta\rho(\rv)=\delta
F[\rho]/\delta\rho(\rv)+V_{\rm ext}(\rv)-\mu$. The intrinsic free
energy density functional $F[\rho]$ contains only adiabatic
contributions, as the many-body distribution at the minimium in
\eqref{EQomegaDensityFunctional} has Boltzmann form
\cite{evans79}. However, as our dynamical theory is built on the
general form $\dot\Omega_{\rm M}[\Psi]$ and thus contains
\eqref{EQomegaMermin}, it retains all non-adiabatic effects.

\subsection{\bf Generating functional}
The many-body function $\hat R$
defined in \eqref{EQgeneratingFunction} contains forces due to
friction, interactions and external fields.  By explicitly separating
off the external field contributions to the total force, $\F^{\rm
  tot}_i(\rnt)$, and substitution of \eqref{EQgeneratingFunction} into
the Levy-type functional \eqref{EQlevy} the {\it intrinsic} part of
the free power functional can be identified as
\begin{align}
  W_t[\rho,\J] =&  {\cal R}_t[\rho,\J]
  +\int d\rv \left(\Xv(\rv,t)-\nabla V_{\rm ext}(\rv,t)\right)\cdot \J(\rv,t)
\notag \\  & 
-\int d\rv \, \dot V_{\rm ext}(\rv,t)\rho(\rv,t)
  \label{EQWintrinsic}
\end{align}
The intrinsic functional retains the same form for all choices of
external field and is therefore universal, in the sense that it only
depends upon the interparticle interactions $U(\rv^N)$. 

As a consequence of the linearity of the external contributions, the 
free power functional acts as a generator for the one-body
density and current, when differentiated with respect to the conjugate
fields,
\begin{align}
  \rho(\rv,t) &= 
  \frac{\delta {\cal R}_t[\rho,\J]}{\delta \dot V_{\rm ext}(\rv,t)},
\label{EQrhoAsDerivative}  \\
  \J(\rv,t) &= -
  \frac{\delta {\cal R}_t[\rho,\J]}{\delta \Xv(\rv,t)},
\label{EQcurrentAsDerivative} 
\end{align}
and hence \eqref{EQWintrinsic} represents a Legendre transform.  For
completeness, the inverse Legendre transform implies that
\begin{align}
  \frac{\delta W_t[\rho,\J]}{\delta \J(\rv,t)}&= \Xv(\rv,t)-\nabla V_{\rm ext}(\rv,t),
  \label{EQinverseLegendreJ}\\
  \frac{\delta W_t[\rho,\J]}{\delta \rho(\rv,t)}&= \alpha(\rv,t)-\dot V_{\rm ext}(\rv,t),
  \label{EQinverseLegendreRho}
\end{align}
where \eqref{EQinverseLegendreJ} follows from the minimization
principle \eqref{EQvariationalPrinciple}, and
\eqref{EQinverseLegendreRho} defines a Lagrange multiplier
$\alpha(\rv,t)=\delta {\cal R}_t[\rho,\J]/\delta \rho(\rv,t)$ to
ensure that the continuity equation can be satisfied. Physically,
$\alpha(\rv,t)$ is a measure of the locally dissipated power.  The
mechanical work created by $V_{\rm ext}(\rv,t)$ is accounted for in
\eqref{EQinverseLegendreJ}, whereas the non-mechanical ``charging''
contribution is contained in \eqref{EQinverseLegendreRho}.  

\subsection{\bf Equation of motion} 
With a view to constructing
approximation schemes it is convenient to split the intrinsic
contribution into a sum of dissipative and reversible contributions
\begin{equation}
  W_t[\rho,\J] =
  P_t[\rho,\J] +
  \int d\rv\,\, \J(\rv,t)\cdot\nabla\frac{\delta F[\rho]}{\delta \rho(\rv,t)},
\end{equation}
where the sum consists of a dissipated power functional,
$P_t[\rho,\J]$, accounting for irreversible energy loss due to the
friction, and a term describing reversible changes in the intrinsic
Helmholtz free energy~\cite{evans79}.  The choice to split the
intrinsic power functional into two terms does not represent a
close-to-equilibrium assumption:
  Nonadiabatic effects are concentrated in the dissipated
power functional. In general, the dissipation functional
$P_t[\rho,\J]$ will be non-local in space and time, depending on the
history of the fields $\rho(\rv,t)$ and $\J(\rv,t)$ prior to time $t$.
Employing the functional chain rule, $\dot F[\rho] = \int \!d\rv\,
\dot\rho(\rv,t) \delta F[\rho]/\delta\rho(\rv,t)$, and the continuity
equation \eqref{EQcontinuityEquation}, followed by a partial
integration in space leads to the alternative expression
\begin{align}
  W_t[\rho,\J]=P_t[\rho,\J]+\dot F[\rho],
  \label{EQgeneralDynamicalFunctional}
\end{align}
where $\dot F[\rho]$ is the total time derivative of the intrinsic
Helmholtz free energy functional. Only the Boltzmann-type
contributions are separated away into $\dot F[\rho]$.  Recalling the
decomposition \eqref{EQRdecomposedIntoMerminDerivative} of the
many-body functional demonstrates that $P_t[\rho,\J]$ contains a
combination of both purely dissipative contributions, via the squared
velocities, and the additional non-adiabatic effects contained within
$\dot \Omega_{\rm M}[\Psi]$, which are in excess of the adiabatic
contribution $\dot F[\rho]$.

Application of the variational principle
\eqref{EQvariationalPrinciple} yields our fundamental equation of
motion
\begin{align}
  \frac{\delta P_t[\rho,\J]}{\delta \J(\rv,t)}=
  -\nabla\frac{\delta F[\rho]}{\delta \rho(\rv,t)}
  -\nabla V_{\rm ext}(\rv,t) + \Xv(\rv,t),
  \label{EQpdft}
\end{align}
where the term on the left hand side of \eqref{EQpdft} represents a
friction force, balanced by the terms on the right hand side arising
from inhomogeneities in the local intrinsic chemical potential,
$\delta F[\rho]/\delta \rho(\rv,t)$, and external forces.  Equation
\eqref{EQpdft} is supplemented by the one-body continuity equation
\eqref{EQcontinuityEquation}.  In equilibrium the left hand side of
\eqref{EQpdft} vanishes and the variational prescription reduces to
\eqref{EQminimizationPrincipleDFT}, thus recovering equilibrium DFT as
a special case. Given that the construction of the dynamical framework
is not based upon the variational principle of DFT
\eqref{EQminimizationPrincipleDFT}, we find this alternative
derivation to be very remarkable, and revealing that the generating
functional ${\cal R}_t[\rho,\J]$, as defined in
\eqref{EQaverage_generator}, is a more fundamental object than the
grand potential density functional $\Omega[\rho]$,
cf.\ \eqref{EQomegaDensityFunctional}.  For tackling non-equilibrium
situations in practice, the challenge is to find explicit forms for
the dissipation functional, $P_t[\rho,\J]$, to obtain a closed
equation of motion.

\subsection{\bf Limiting cases}
In order to gain some intuition into the
equation of motion \eqref{EQpdft}, we consider three special limiting
cases:

{\it (i) Instantaneous motion}. The mathematically simplest case is
obtained by neglecting nonconservative external forces and setting
$P_t[\rho,\J]=0$.  Equation \eqref{EQpdft} then yields $\delta
F[\rho]/\delta \rho(\rv,t)=\mu-V_{\rm ext}(\rv,t)$, which is identical
to the Euler-Lagrange equation of equilibrium DFT, where the chemical
potential, $\mu$, keeps the particle number constant.  Hence the
density field instantaneously follows changes in the external
potential.

{\it (ii) Ideal gas}. For a system of noninteracting particles the
Helmholtz free energy is known exactly, $F_{\rm id}[\rho]=\kT\int
d\rv \rho(\rv,t) [\,\ln(\rho(\rv,t)\Lambda^3)-1]$, where $\Lambda$ is
the thermal wavelength, and the dissipation functional is given by
\begin{align}
  P_t^{\rm id}[\rho,\J]=\int d\rv\frac{\gamma
    \J(\rv,t)^2}{2\rho(\rv,t)}.
  \label{EQpid}
\end{align}
Functional differentiation of this expression at fixed time with
respect to the current generates the (mean) friction force, such that
$\J(\rv,t)\cdot\delta P_t^{\rm id}[\rho,\J]/\delta
\J(\rv,t)=\J(\rv,t)\cdot\gamma \vel(\rv,t)$ represents the dissipated
power density resulting from the average motion of the system; here
$\vel(\rv,t)=\J(\rv,t)/\rho(\rv,t)$ is the average one-body velocity.
The exact expression for the ideal current, $\gamma\J_{\rm
  id}(\rv,t)/\rho(\rv,t) = -\kT\nabla\ln\rho(\rv,t)-\nabla V_{\rm
  ext}(\rv,t)+\Xv(\rv,t)$ follows from substitution of \eqref{EQpid}
into \eqref{EQpdft}.

{\it (iii) Dynamical density functional theory}. The intrinsic free
energy functional of an interacting system can be written as the sum
of two contributions, $F[\rho]=F_{\rm id}[\rho]+ F_{\rm exc}[\rho]$,
where the excess contribution accounts for interparticle interactions.
If we retain the full free energy, but assume that the dissipation is
given by \eqref{EQpid}, then \eqref{EQpdft} yields $\J(\rv,t)=\J_{\rm
  DDFT}(\rv,t)$, where
\begin{align}
  \hspace*{-0.08cm}\frac{\gamma\J_{\rm DDFT}(\rv,t)}{\rho(\rv,t)} =
    -\nabla \frac{\delta F[\rho]}{\delta \rho(\rv,t)}
    -\nabla V_{\rm ext}(\rv,t) + \Xv(\rv,t)
    \label{EQjddft}
\end{align}
is precisely the current of DDFT
\cite{evans79,marinibettolomarconi99}.  We thus gain new insight into
the standard theory, namely that the adiabatic approximation is
equivalent to assuming a trivial, noninteracting form for the
dissipation functional.  This observation suggests that superior
theories can be obtained by developing approximations to
$P_t[\rho,\J]$ which recognize the existence of interparticle
interactions.  

\subsection{\bf Beyond DDFT} 
We can now go beyond DDFT by decomposing
the dissipation power functional into two contributions,
\begin{align}
  P_t[\rho,\J]=P_t^{\rm id}[\rho,\J]+P_t^{\rm exc}[\rho,\J],
  \label{EQPsplitting}
\end{align}
where $P_t^{\rm id}[\rho,\J]$ is given by \eqref{EQpid} and $P_t^{\rm
  exc}[\rho,\J]$ accounts for dissipation arising from interparticle
interactions. Equation \eqref{EQPsplitting} is to be viewed as the
definition of $P_t^{\rm exc}[\rho,\J]$ rather than as an assumption.
Microscopically, the excess dissipation is caused by the non-adiabatic
contributions to the time derivative of the Mermin functional
\eqref{EQomegaMermin}, and hence $P_{\rm exc}[\rho,\J]$ depends
explicitly on the interparticle interactions $U(\rv^N)$.  Using
\eqref{EQPsplitting}, the fundamental equation of motion
\eqref{EQpdft} becomes
\begin{align}
  \frac{\gamma \J(\rv,t)}{\rho(\rv,t)} &+
  \frac{\delta P_t^{\rm exc}[\rho,\J]}{\delta \J(\rv,t)}=
  \notag\\
  & -\nabla\frac{\delta F[\rho]}{\delta \rho(\rv,t)}
  -\nabla V_{\rm ext}(\rv,t) + \Xv(\rv,t).
  \label{EQpdftPexc}
\end{align}
Corrections to DDFT thus come from the second term on the left hand
side. It is instructive to compare \eqref{EQpdftPexc} with the
formally exact result obtained by integrating the many-body
Smoluchowski equation, given by \eqref{EQsmol} and
\eqref{EQsmol_current}, over $N-1$ particle coordinates
\cite{archer04ddft}.  Restricting ourselves for simplicity to systems
that interact via pair potentials, $U=\sum_{i<
  j}\phi(\rv_i,\rv_j)$, an exact equation of motion is obtained,
\begin{align}
  \frac{\gamma\J(\rv,t)}{\rho(\rv,t)}=&
  -\kT \nabla \ln(\rho(\rv,t)\Lambda^3)-\nabla V_{\rm ext}(\rv,t)
  \nonumber\\&
  +\Xv(\rv,t)
  -\int d\rv'\frac{\rho^{(2)}(\rv,\rv',t)}{\rho(\rv,t)}\nabla\phi(\rv,\rv'),
\end{align}
where $\rho^{(2)}(\rv,\rv',t)$ is the exact (but unknown) out-of-equilibrium
equal-time two-body density. Comparison with \eqref{EQpdftPexc} yields
\begin{align}
  \!\!\!\frac{\delta P_t^{\rm exc}[\rho,\J]}{\delta \J(\rv,t)} &= 
  -\nabla \frac{\delta F_{\rm exc}[\rho]}{\delta \rho(\rv,t)}
  +\int d\rv' \frac{\rho^{(2)}(\rv,\rv',t)}{\rho(\rv,t)}
  \nabla\phi(\rv,\rv').
\end{align}
Splitting the full non-equilibrium two-body density into an
instantaneous-equilibrium and an irreducible part,
$\rho^{(2)}(\rv,\rv',t)=\rho^{(2)}_{\rm
  eq}(\rv,\rv',t)+\rho^{(2)}_{\rm irr}(\rv,\rv',t)$, leads to the
following pair of identities
\begin{align}
  \nabla \frac{\delta F_{\rm exc}[\rho]}{\delta \rho(\rv,t)}
  &=\int d\rv' \frac{\rho^{(2)}_{\rm eq}(\rv,\rv',t)}{\rho(\rv,t)}
  \nabla\phi(\rv,\rv'),
  \label{EQsumRuleEquilibrium}\\
  \frac{\delta P_t^{\rm exc}[\rho,\J]}{\delta \J(\rv,t)}
  &=\int d\rv' \frac{\rho^{(2)}_{\rm irr}(\rv,\rv',t)}{\rho(\rv,t)}
  \nabla\phi(\rv,\rv').
  \label{EQsumRuleNonEquilibrium}
\end{align}
Equation~\eqref{EQsumRuleEquilibrium} is an exact equilibrium relation
\cite{archer04ddft}, whereas \eqref{EQsumRuleNonEquilibrium} expresses
the fact that the excess dissipation is intimately connected with
beyond-adiabatic, irreducible two-point correlations. This is fully
consistent with the interpretation that the dynamic functional is a
generator for many-body induced friction forces in the system.  If the
system contains many-body forces, generalized versions of the
identities \eqref{EQsumRuleEquilibrium} and
\eqref{EQsumRuleNonEquilibrium} hold, which contain three- and
higher-body density distributions.
\begin{figure}[t!]
\hspace*{0.cm}
\includegraphics[width=8cm,angle=0]{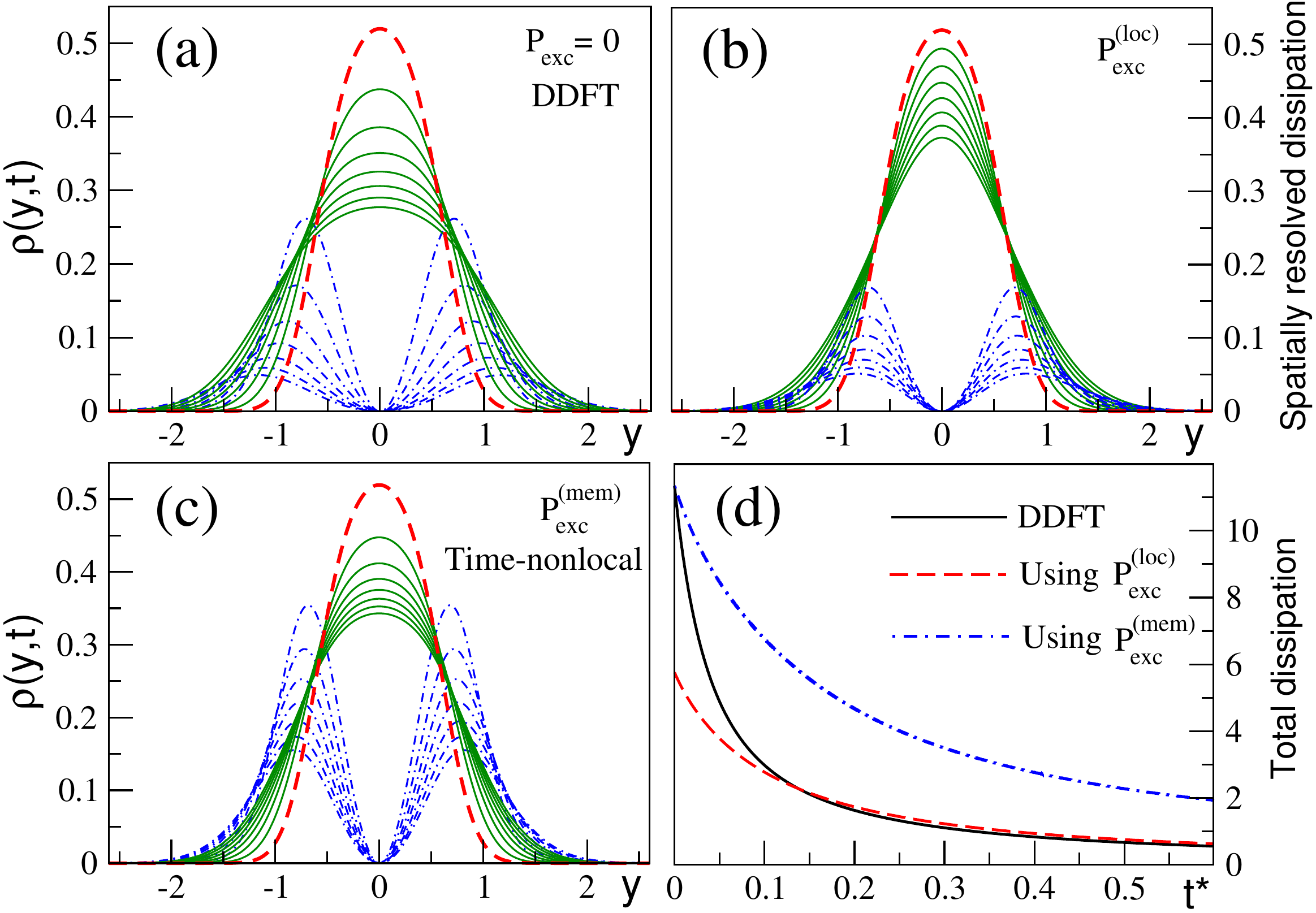}
\caption{Relaxation of a density peak (slab geometry) from an
  initial state (red dashed) for which the density is confined
  to a parabolic potential. Solid green curves show the density
  and blue dash-dotted curves show the spatially resolved
  dissipation (scaled down by a factor 20) for three approximations 
  to the excess dissipation (at dimensionless times
  $t^*\!\equiv t\kT/\gamma d^2\!\!=\!0.2$ to $1.4$ in steps of $0.2$).
  Using (a) DDFT, (b) \eqref{EQrescale}, (c) \eqref{EQnonlocal_time}. 
  (d) Decay of the total dissipated power.}
\label{relaxation}
\end{figure}

\subsection{\bf Approximations to the excess dissipation} 
Any practical application of the framework, i.e.\ implementation of
\eqref{EQpdftPexc} along with \eqref{EQcontinuityEquation}, requires
one to choose an approximation for $P_t^{\rm exc}[\rho,\J]$ for the
given physical system (as characterized by $U$). Here we discuss
several physically reasonable approximations to $P_t^{\rm
  exc}[\rho,\J]$, which may be applied to treat systems with arbitrary
pairwise interactions.  The simplest of these is given by the space-
and time-local expression
\begin{align}
  P_t^{\rm loc}[\rho,\J] = \int d\rv\, 
  \left(\tilde\gamma(\rho)-\gamma\right)\frac{\J(\rv,t)^2}{2\rho(\rv,t)},
\label{EQrescale}
\end{align}
where $\tilde\gamma(\rho)$ is a density-dependent friction factor
which can be approximated using either a virial expansion or, in the
case of hard spheres, by employing closed form expressions
\cite{hayakawa95}.  Using \eqref{EQrescale} slows the dynamics in
regions of high density relative to DDFT \cite{royallSedimentation},
but will still fail to describe collision induced dissipation arising
from shearing motions in the fluid.  To capture this requires a
nonlocal treatment in space and in time which recognizes the finite
range of the colloidal interactions. A suitably general form is given
by
\begin{align}
  \!\!\!\!P_t^{\rm nloc}[\rho,\J]=\!\!
  \int\!\! d\rv\!\! \int\!\! d\rv'\!\!\! \int_{t_0}^t \!\!\! dt'\,
  \J(\rv,t)\cdot {\bf K}(\rv,t;\rv',t') \cdot\J(\rv',t'),
\label{EQnonlocal}
\end{align}
where the convolution kernel ${\bf K}$ is a second-rank tensor
describing current-current scattering and is in general a functional
of $\rho$.  The incorporation of temporal nonlocality represents a
powerful feature of the present approach.  A simple specific form for
the scattering kernel which captures this is ${\bf K}=\gamma
m(t-t')\delta(\rv-\rv')/\rho(\rv,t)$, where $m(t-t')$ is a memory
function (with units of inverse time). As a result we obtain a
spatially local excess dissipation functional with temporal memory,
given by
\begin{align} 
  P_t^{\rm mem}[\rho,\J] = \int d\rv
  \int_{t_0}^t dt' 
  \frac{\gamma \J(\rv,t)\cdot\J(\rv,t')}{\rho(\rv,t)} m(t-t').
\label{EQnonlocal_time}
\end{align}
For general interactions and time-dependent external fields the memory
need not be time translationally invariant, as assumed here. 
The equation of motion \eqref{EQpdftPexc} thus becomes
\begin{align}
  \frac{\gamma }{\rho(\rv,t)}&\left( \J(\rv,t) +
  \int_{t_0}^t dt' \J(\rv,t') m(t-t') \right) = \notag\\
  & -\nabla\frac{\delta F[\rho]}{\delta \rho(\rv,t)}
  -\nabla V_{\rm ext}(\rv,t) + \Xv(\rv,t),
\end{align}
which captures the history-dependence of the one-body current, but
neglects the spatial nonlocality of the dissipation.  

\subsection{\bf Memory effects: A numerical test} 
We have performed
numerical calculations using the (three-dimensional) hard sphere
system in a slab geometry using approximations \eqref{EQrescale} and
\eqref{EQnonlocal_time}.  The (spatially non-local) Rosenfeld
functional \cite{Rosenfeld89} was used to approximate $F_{\rm
  exc}[\rho]$.  The system is initially confined along the $y$-axis by
a parabolic potential, $V_{\rm ext}(\rv,t<0)=b y^2/2$, where $b$ is a
constant measuring the strength of the trapping potential. We then
switch off the confinement at $t=t_0=0$, such that $V_{\rm
  ext}(\rv,t>0)=0$, and monitor the time evolution of the density.  In
Fig.1 we show the density calculated using three approximations: DDFT,
the density-dependent friction coefficient \eqref{EQrescale} (where we
have employed the expression for $\tilde{\gamma}(\rho)$ from
\cite{hayakawa95}, and the temporally nonlocal approximation
\eqref{EQnonlocal_time}.  When implementing \eqref{EQnonlocal_time} we
have approximated the memory function by the simple form
$m(t-t')=(a/\tau)\exp(-(t-t')/\tau)$, where $a$ is a dimensionless
parameter and $\tau$ is a relaxation time.  In view of the complexity
of mode-coupling-type memory functions \cite{sjoegren} it is clear
that the assumption of exponential decay represents a strong
simplification, but nevertheless provides a first step
towards recognizing the history dependence of the one-body fields.

The density profiles in Fig.1 (generated using the parameter values
$a\!=\!0.8, b\!=\!10\kT, \tau\!=\!0.05\gamma d^2/\kT$, where the
particle diameter, $d$, sets the length-scale) show that both
\eqref{EQrescale} and \eqref{EQnonlocal_time} slow the dynamics.  The
magnitude of the retardation achieved using \eqref{EQnonlocal_time}
depends upon the choice of parameters $a$ and $\tau$ determining the
memory.  While \eqref{EQnonlocal_time} generates density profiles with
a similar functional form to those from DDFT, Eq.~\eqref{EQrescale}
leads to a density more sharply peaked at the origin. The spatially
resolved dissipation (blue dash-dotted curves) confirms the intuition
that power is dissipated mostly in regions of high density gradient
and the spatial integral of this quantity, $\int \!d\rv\,
\J(\rv,t)\cdot\delta P_t[\rho,\J]/\delta \J(\rv,t)$, decays towards
zero as the system approaches equilibrium.  It is well known that
relaxation rates predicted by standard DDFT are significantly faster
than those found in simulation
\cite{marinibettolomarconi99,royallSedimentation}. From 
our findings we conclude that this failing can be remedied by the
incorporation of temporal nonlocality in the excess dissipation
functional, leading to memory effects in the equation of motion for
$\J(\rv,t)$.  If an exponential memory is employed then the
computational demands are comparable to those of DDFT.  

\section{\bf Conclusions} 
In summary, we have shown that collective
Brownian dynamics can be formulated as a variational theory
\eqref{EQvariationalPrinciple} based upon the dissipative power as a
functional of the one-body density and the one-body current.  The
underlying many-body expression \eqref{EQaverage_generator} is a
difference of half of the power that is dissipated due to friction and
the total power that is generated by the deterministic and entropic
forces. As we have shown, this free power functional is minimal for the
physical time evolution, and hence plays a role analogous to that of
the free energy functional in equilibrium. Thermodynamic potentials in
equilibrium are abstract quantities, detectable only through their
derivatives. The same is true for the power functional, cf.\ the
equation of motion \eqref{EQpdft}, which hence attains a similar
status, but is of more general nature, as it applies out of equilibrium.
We have formulated the theory in the Smoluchowski picture, starting
with the time evolution of the many-body probability distribution
$\Psi(\rnt)$. An alternative derivation could be based upon the
Langevin equation \eqref{EQlangevin}, using the path integral approach to
obtain averaged quantities \cite{onsager}. In both cases, the one-body
current and density are averages that do not fluctuate,
despite any formal similarities of our approach to dynamical field
theories, where the fields themselves can fluctuate.

The appeal of our approach stems from its utility and and ease with
which it can be implemented, which contrast strongly with the
time-dependent classical DFT formalism by Chan and Finken
\cite{chan05}. As far as we are aware, the approach of \cite{chan05}
has never been applied to any model system, possibly because it is
built around an action functional, which so far could not be
approximated in any systematic or physically intuitive way. This is
not the case with the excess dissipation functional identified in the
present work, for which even simple expressions, such as
\eqref{EQnonlocal_time}, transcend the adiabatic approximation. While
DDFT remains an active field of research (e.g.\ hydrodynamic
interactions \cite{serafim12} and arbitrary particle shapes
\cite{wittkowski12} have very recently been addressed), it is
important to appreciate that all extensions and modifications proposed
since the original presentation of DDFT \cite{evans79} have been
firmly under the constraint of adiabaticity. The power functional
approach is free of this restriction and provides a solid, nonadiabatic  
basis for extensions aiming to treat more complex model systems 
(e.g. orientational degrees of freedom). Moreover, it applies also to systems 
governed by many-body forces, where $U(\rv^N)$ also contains three- and
higher-body contributions. 
The only other nonadiabatic approach of which we are aware is the 
Generalized Langevin theory of M.~Medino-Noyola and coworkers \cite{magdeleno}. 
Exploring connections to this work may prove fruitful.  

Regarding extensions of the power functional theory: Firstly, generalization 
to mixtures of different species is straightforward. The ideal 
contribution becomes $\sum_i P_t^{\rm id}[\rho_i,\J_i]$, where $i$ enumerates the different
species, whereas the excess dissipation, $P_t^{\rm exc}[\{\rho_i\},\{\J_i\}]$, generates
dynamical coupling between particles of different species. 
Secondly, if the Langevin equation \eqref{EQlangevin} is generalized to
include a velocity-dependent friction coefficient,
$\gamma(\vel_i(t))$, a modified version of the ideal dissipation
functional applies, $P_t^{\rm id, nl}[\rho,\J]=\int d\rv
\rho(\rv,t)f(\vel(\rv,t)^2)$, where the function $f(\cdot)$ is related
to the density-dependent friction force via $\gamma(\vel^2)=2\rho
f'(\vel^2)$ and the prime denotes differentiation with respect to the
argument.

Much of the phenomenology of non-equilibrium dynamics can be
investigated using two- and higher-body correlation functions. Using
the dynamical test particle method \cite{archer07dtpl}, which
identifies the van Hove function with the dynamics of suitably
constructed one-body fields, one has immediate access within the
present framework to the dynamic structure factor and the intermediate
scattering function. Moreover, the relationships
\eqref{EQrhoAsDerivative} and \eqref{EQcurrentAsDerivative} of the
one-body fields to their generating functional imply that two- and
higher-body dynamic correlation functions can be generated from
further functional differentiation, putting a non-equilibrium
generalization of the Ornstein-Zernike relation, which in equilibrium
is a cornerstone of liquid state theory \cite{hansen}, within reach.
Work along these lines, as well as application to driven lattice
models \cite{dwandaru}, is currently in progress.

\begin{acknowledgments}
We thank 
A.~J. Archer,
M.~E. Cates,
S. Dietrich,
D. de las Heras,
R. Evans,
T.~M.~Fischer,
M. Fuchs,
B. Goddard,
S. Kalliadasis,
and F. Schmid
for useful discussions and comments. 
\end{acknowledgments}

\end{document}